\documentclass[final, 3p, a4paper]{elsarticle}

\usepackage{papersettings}
\usepackage{mathsettings}
\usetikzlibrary{fit}
\usetikzlibrary{angles}
\usepackage{setspace}
\newtheorem{remark}{Remark}

\modulolinenumbers[1]

\graphicspath{{./figures/}}

\journal{}

\begin{document}

\begin{frontmatter}
	% \title{A fully Eulerian phase-field model for frictional contact between evolving solids}
    % \title{An implicit Eulerian phase-field model for frictional contact between elastic solids}
    % \title{Frictional contact in a Eulerian phase-field framework}
    \title{Frictional contact between solids: A fully Eulerian phase-field approach}

	\address[ifbaddress]{Institute for Building Materials, ETH Zurich, Switzerland}
	\author[ifbaddress]{Flavio Lorez}
	\ead{florez@ethz.ch}
	\author[ifbaddress]{Mohit Pundir\corref{cor1}}   \ead{mpundir@ethz.ch}
	% \author[ifbaddress]{David S. Kammer \corref{cor1}} \ead{dkammer@ethz.ch}
	\cortext[cor1]{Corresponding author}

	\begin{abstract}
        Recent advancements have demonstrated that fully Eulerian methods can effectively model
        frictionless contact between deformable solids. Unlike traditional Lagrangian approaches, which
        require contact detection and resolution algorithms, the Eulerian framework utilizes a single,
        fixed spatial mesh combined with a diffuse interface phase-field approach, simplifying
        contact resolution significantly. Moreover, the Eulerian method is well-suited for
        developing a unified framework to handle multiphysical systems involving growing bodies that
        interact with a constraining medium. 
        In this work, we extend our previous methodology to incorporate frictional contact. By
        leveraging the intersection of the phase fields of multiple bodies, we define normal and
        tangential penalty force fields, which are incorporated into the linear momentum equations
        to capture frictional interactions. This formulation allows independent motion of each body
        using distinct velocity fields, coupled solely through interfacial forces arising from contact and friction.
        We thoroughly validate the proposed approach through several numerical examples. The method
        is shown to handle large sliding effortlessly, accurately capture the stick-slip transition,
        and preserve history-dependent energy dissipation, offering a solution for modeling
        frictional contact in Eulerian models.

	\end{abstract}


\end{frontmatter}

% \linenumbers

%\section*{Highlights}
%\begin{itemize}
%    \item Novel Eulerian phase-field framework for frictional contact between elastic solids.
%    \item First methodology to integrate dry friction within a fully Eulerian context.
%    \item Solids are modeled by a combination of the phase-field method and the reference-map technique.
%    \item Systematic verification against the analytical solution of the Cattaneo-Mindlin problem.
%    \item Demonstration of history-dependency, large sliding, and multi-patch contact.
%\end{itemize}

\section{Introduction}
\label{sec:introduction}

Interfacial interactions are not only fundamental in various mechanical systems, ranging from
engineering materials to natural phenomena, but also play a pivotal role in multiphysical processes,
where they often arise as a consequence of other mechanisms and, in turn, influence these underlying
processes.
For instance, in biofilms, frictional contact and adhesion between
growing bacterial colonies play a crucial role in the emergence of complex pattern formations and
mechanical stability~\cite{amar_patterns_2014}. Similarly, frictional interactions play a significant role
in the corrosion-driven failure of cementitious materials driven by precipitate growth~\cite{angst_challenges_2018,pundir_fft-based_2023}. Capturing these interactions
accurately within computational models is notoriously difficult due to the intricate interplay
between contact forces and evolving geometries.
In solid mechanics, contact problems are predominantly modeled using Lagrangian frameworks because
the direct association between material points and discretization nodes facilitates straightforward
tracking of displacements from the reference configuration, which is essential for implementing
elastic constitutive laws. While the body-specific domains necessitate contact detection and
resolution algorithms to couple different bodies~\cite{yastrebov_computational_2012}, this close coupling makes the Lagrangian approach
inherently well-suited for modeling the behavior of solid materials. 
Unfortunately, the Lagrangian description is not well suited for modeling evolving interfaces, such as in
physical, chemical, or biological growth processes. 
A significant limitation is that any surface growth within a body requires a re-meshing procedure to
accommodate changes in geometry. Nonetheless, such adaptations have been successfully implemented in
certain cases (\eg,~\cite{li_nonlinear_2022}). Moreover,
handling contact between evolving surfaces using Lagrangian finite elements is particularly
costly due to the need for continuous detection and resolution of new contact
regions. Consequently, existing Lagrangian approaches for
modeling growth in constrained space simplify interfacial interactions by assuming perfect bonding
and neglect complex phenomena like frictional sliding and adhesion, often relying on approximations
such as Eshelby's inclusion theory~\cite{eshelby_elastic_1997}.

In contrast, Eulerian methods, where the computational mesh is fixed in space, provide a more straightforward
approach to modeling evolving interfaces, being the natural choice for diffusion, transport or
growth processes~\cite{soleimani_numerical_2023}. 
Although incorporating solid mechanics in Eulerian frameworks involves
challenges such as surface tracking and obtaining the total deformation, these
obstacles can be effectively addressed using established methodologies. Common
approaches for tracking solid boundaries include the level-set
method~\cite{osher_level_2003} and the phase-field
technique~\cite{sun_sharp_2007,boettinger_phase-field_2002}. The computation of the deformation
gradient, a key requirement for solid mechanics, can be achieved through the
reference map technique, which enables precise calculation of total deformation
within a fixed Eulerian
mesh~\cite{kamrin_reference_2012,rycroft_reference_2020,daubner_multiphase-field_2023}. These advancements
position the Eulerian approach as a promising option for unified modeling across
a variety of physical processes.

In recent years, several studies, primarily in the context of fluid-structure interaction, have
demonstrated that enforcing continuity of a single velocity field across the entire computational
domain can effectively transmit pressure between contacting
bodies~\cite{valkov_eulerian_2015,rycroft_reference_2020,mao_3d_2024,rath_efficient_2024}. However,
this approach inherently maintains a persistent gap between the bodies and restricts arbitrary
sliding at the interface. To address these limitations, we recently introduced a new methodology for
modeling frictionless contact within a fully Eulerian framework, where each body is represented
separately using distinct motion functions and field variables~\cite{lorez_eulerian_2024}. By
combining the reference map technique with the phase-field method to represent multiple bodies, we
modeled contact using an implicit function defined in terms of the phase-field variables.
This formulation enables arbitrary sliding between bodies and effectively handles large deformations
and complex geometries without the need for explicit contact tracking. By separating interface
interactions from the bulk motion of the bodies, we proposed a flexible framework that can be adapted
to model more complex interfacial phenomena.
Building on these foundations, we introduce a novel Eulerian formulation that systematically
integrates frictional contact into the modeling framework. By employing separate sets of Eulerian
field variables for each solid, similar to a Lagrangian multi-body representation, we preserve the
independence of each body while enabling efficient coupling through contact constraints. This
approach leverages the strengths of Eulerian methods such as simplified contact detection on a
shared mesh and overcomes their traditional limitation of spurious bonding, thereby enabling
accurate handling of frictional slip and stick behavior along interfaces.

The remainder of the paper is organized as follows. In Section~\ref{sec:method}, we present the
Eulerian description of soft body motion and the incorporation of elasticity and the phase-field
method for interface capturing. Section~\ref{sec:contact-formulation} revisits the frictionless
contact formulation from~\cite{lorez_eulerian_2024} and details the frictional contact
formulation using penalty-based body forces. The numerical implementation is discussed in
Section~\ref{sec:numerical_implementation}. In Section~\ref{sec:examples}, we provide numerical
examples to validate and demonstrate the capabilities of the method. Finally, we conclude the paper
with a discussion of advantages, challenges, and potential future work in
Section~\ref{sec:discussion}.

% --------------------------------------------------------------------------------------------

\section{Eulerian description of the motion of elastic bodies}
\label{sec:method}

In a classical Lagrangian description, the evolution of an elastic body $\mathcal{B}$ is
sufficiently described by the deformation map $\boldsymbol x = \chi(\X, t) : \mathbb{R}^D \times [0,
T] \rightarrow \mathbb{R}^D$ ($\x$ being the Eulerian coordinates, $\X$ being the Lagrangian
reference coordinates), that transforms the reference configuration $\mathcal{B}_0$ into the current
configuration $\mathcal{B}(t)$ for all times $t$. The tangent of $\chi$, the \textit{deformation
gradient} $\boldsymbol{F} = \partial \x/ \partial \X$ describes the strain state. 

\subsection{Eulerian description of elasticity}

To model elastic solids in an Eulerian frame, we employ the reference map
technique~\cite{kamrin_reference_2012,dunne_eulerian_2006}, which introduces the reference map
$\X = \xi(\x, t) : \Omega \times [0, T] \rightarrow \mathbb{R}^D$, that maps the reference to the current
configuration (see \cref{fig:methodology}a). This approach allows for a consistent approximation of the deformation gradient, and
consequently a way to model elastic solid laws in Eulerian frameworks. The deformation gradient is the inverse of the gradient of the reference map:
\begin{equation}
    \dfrac{\partial\xi}{\partial\x} = \nabla\xi ~, \quad
    \boldsymbol{F}(\x, t) = \dfrac{\partial\x}{\partial\X} = \left(\nabla\xi(\x, t)\right)^{-1} ~.
	\label{eq:deformation_gradient}
\end{equation}

Given the material velocity $\vv$, any Eulerian field must be transported accordingly. For the reference position to remain unchanged from the perspective of a material point, the material time derivative $D\xi/Dt = 0$, provides the evolution
law for the reference map:
\begin{equation}
    \dfrac{D\xi}{Dt} = \dfrac{\partial\xi}{\partial t} + \vv \cdot \nabla\xi = 0 ~.
    \label{eq:reference-map-evolution}   
\end{equation}

With access to the deformation gradient $\boldsymbol{F}$ through the reference map $\xi$, any elastic constitutive relation can be formulated, such that:
\begin{equation}
    \boldsymbol{\sigma} = f(\boldsymbol{F}) ~,
\end{equation}
where $\boldsymbol{\sigma}$ represents the stress tensor, and $f(\boldsymbol{F})$ describes the elastic material behavior.

The specific constitutive formulations utilized in this work are provided within the respective examples in \cref{sec:examples}.

\subsection{Interface capturing with the phase-field method}
\label{sec:phase-field-method}

\begin{figure}
    \begin{center}
        \includegraphics{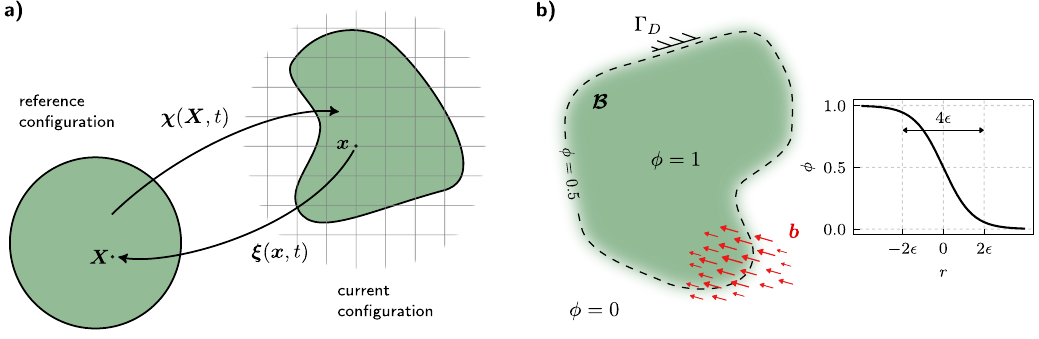}
    \end{center}
    \caption{
    \textbf{(a)} An illustration of the reference map $\xi$. The reference map $\xi(\x, t)$ -- being the inverse motion function $\chi(\X, t)$ -- returns the reference location $\X$ for any Eulerian coordinate $\x$.
    \textbf{(b)} Illustration of a body $\mathcal{B}$ implicitly defined by a diffuse phase-field. The smooth transition of its boundary from $\phi=1$ to $\phi=0$ is shown in a 1D cut. The width of the transition zone is characterised by the parameter $\epsilon$, where $\sim 90\%$ of the variation in $\phi$ lies within a width of $4\epsilon$. The possible presence of Dirichlet boundary conditions and force fields is indicated with $\Gamma_D$ and $\boldsymbol{b}$.
    }
    \label{fig:methodology}
\end{figure}

We adopt the phase-field method to implicitly capture the domain of each body $\mathcal{B}_i$ and
its boundary $\partial \mathcal{B}_i$, by introducing auxiliary scalar fields $\phi_i(\x, t) \in
[0, 1]$, where $\phi_i = 1$ and $\phi_i = 0$ define its solid and void phases, respectively (see \cref{fig:methodology}b). 
The phase-field method is a diffuse interface approximation, characterized by a smooth transition
over a finite but small width from solid to void phase that originates from the phase-field free
energy functional~\cite{cahn_free_1958,sun_sharp_2007}:
\begin{equation}
    \mathcal{E}_{\text{PF}}(\phi) = \int_\Omega \left( \dfrac{\epsilon^2}{2} (\nabla\phi)^2 + g(\phi) \right) ~\mathrm{d}\Omega ~,
    \label{eq:phase-field-energy}
\end{equation}
where $\epsilon^2$ is the energy gradient coefficient, and $g(\phi) = \phi^2(1-\phi)^2$ is a
double-well potential with minima located at 0 and 1.
The energetically favorable state of \cref{eq:phase-field-energy} yields the well-known tangent-hyperbolicus
equilibrium profile normal to the interface, as shown in the inset of \cref{fig:methodology}b,
\begin{equation}
    \phi(r) = \dfrac{1}{2} \left[1 + \tanh\left(\dfrac{r}{\sqrt{2}\epsilon}\right)\right] ~,
    \label{eq:interface-profile}
\end{equation}
where $r$ is the signed distance from the sharp interface.

In the absence of phase transformation and under pure mechanical deformation, the
phase-field parameter $\phi_i$ should remain constant with respect to the reference coordinates
$\X_i$.
Equivalent to \cref{eq:reference-map-evolution}, the material time derivative of $\phi_i$ yields the
general evolution law
\begin{equation}
	\dfrac{d\phi}{dt} = \dfrac{\partial\phi}{\partial t} + \boldsymbol v \cdot \nabla\phi = 0 ~.
	\label{eq:phase-field-advection}
\end{equation}

While \cref{eq:phase-field-advection} theoretically preserves the original phase for all material points, it leads
to a distortion of the diffuse interface, a phenomenon referred to as advective distortion~\cite{daubner_multiphase-field_2023,sun_sharp_2007,mao_variational_2021,mao_interface_2023}. To
counteract this, we consider the gradient flow towards minimizing the phase-field energy using an
advective Cahn-Hilliard equation~\cite{li_solving_2009,cahn_free_1958,aland_phase_2017}:
\begin{equation}
    \dfrac{d\phi}{dt} = \dfrac{\partial\phi}{\partial t} + \vv\cdot\nabla\phi =
    \nabla\cdot\left(-\mathcal{M}\nabla q\right)~, \quad
    q = \dfrac{\delta\mathcal{E}_{\text{PF}}}{\delta\phi}
	\label{eq:cahn-hilliard}
\end{equation}
where $\mathcal{M}$ is a mobility coefficient controlling the magnitude of regularization. It must
be highlighted that this regularization introduces an error in the conservation of the original
body and must, therefore, be chosen carefully.
To mitigate advective distortion while minimizing the impact of regularization, time-dependent mobility models have been proposed~\cite{mao_variational_2021}.
However, for simplicity, we adopt a constant mobility $\mathcal{M}$ in this work.

We have chosen the fourth-order Cahn-Hilliard equation over the second-order Allen-Cahn equation because it is naturally volume
conserving. Previous authors have presented concepts to make the Allen-Cahn equation volume conserving,
\eg~\cite{mao_variational_2021,mao_interface_2023,mao_3d_2024,aland_phase_2017}. However, the mass correction term is
usually non-local which complicates our implementation. As we will present in \cref{sec:numerical_implementation},
we construct a single non-staggered variational minimization problem including the linear momentum
balance, the phase-field and reference-map evolution equations. For similar reasons, we
have refrained from making the phase-field a \textit{constant} function of the reference map, as it has been
done with level-set~\cite{rycroft_reference_2020} or phase-field
functions~\cite{daubner_multiphase-field_2023}. We require the phase-field to be part of the
solution space because we later need its variation to find the optimal contact forces.

\subsection{Conservation of linear momentum}

The phase-fields $\phi$ track the interfaces while the reference maps $\xi$ track the motion of material
points. The combination of these two fields allows a
definite description of the motion of deformable solid bodies in an Eulerian framework.

At time $t$, given some previous configuration $\mathcal{B}_t$ with field variables $\phi_t$ and $\xi_t$, we aim to
find a new configuration $\mathcal{B}_{t+1}$ with field variables $\phi_{t+1}$ and $\xi_{t+1}$ such
that the linear momentum balance is satisfied:
\begin{equation}
    \nabla \cdot \left(\phi\boldsymbol{\sigma}\right) - \boldsymbol{b} = \boldsymbol{0} ~,
	\label{eq:equilibrium}
\end{equation}
where $\boldsymbol{b}$ is an external force density. Contact forces are included as external forces,
as will be described in \cref{sec:contact-formulation}.

% POSSIBLY MOVE LATER
\begin{remark}
    Reviewing \cref{eq:reference-map-evolution}, the velocity $\vv$ and reference map $\xi$ can be strongly coupled.
    Therefore, we define the velocity field $\vv(\xi)$ as an explicit function of the reference map $\xi$, effectively decreasing the number of unknowns:
\begin{equation}
	\boldsymbol{v}(\xi, t) = -\left( \nabla\xi\right)^{-1} \cdot \dfrac{\partial\xi}{\partial t}~.
	\label{eq:velocity_field}
\end{equation}
Because the weak form, which we will introduce in \cref{sec:numerical_implementation}, has no terms
containing $\partial\vv/\partial x$, the velocity field $\vv$ is not required to be continuous. This
allows for a discontinuous velocity field, obtained from the continuous space used for $\xi$.
\end{remark}

%  ----------------------------------------------------------------------------------------

\section{Frictional contact formulation}
\label{sec:contact-formulation}

\begin{figure}
    \begin{center}
        \includegraphics{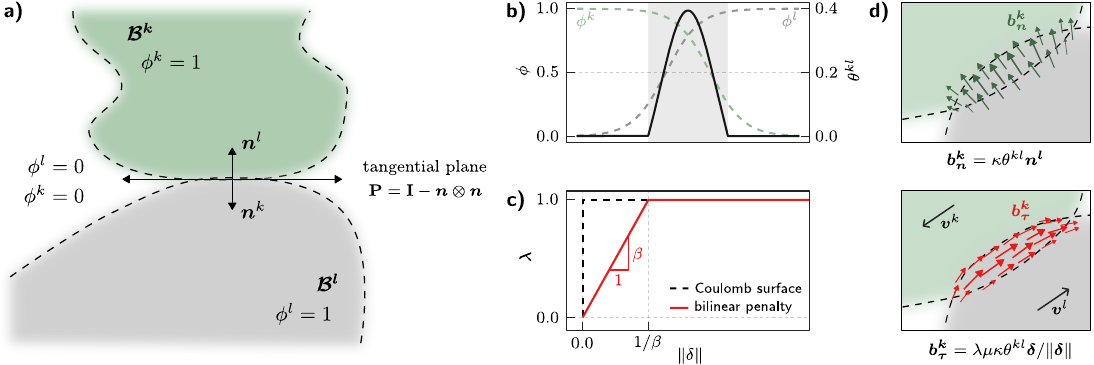}
    \end{center}
    \caption{Schematic of the frictional contact formulation. 
        \textbf{(a)} The two phase-fields $\phi_k$ and $\phi_l$ represent two elastic bodies in
        contact. We consider them to be in contact if their $0.5$ iso-level sets intersect. The
        normals are used to define a projection matrix $\mathbf{P}$ that is used to project the
        relative velocity $\overline{\vv}_k$ onto the tangential plane.
       \textbf{(b)} 1D cut normal to the surfaces of the overlapping region. The intersection
        variable $\theta_{kl}$ is non-zero only in the immediate neighbourhood of regions where both
        phase-fields are $>0.5$.
        \textbf{(c)} The frictional penalty function, which is used to approximate the Coulomb friction law.
        \textbf{(d)} Illustration of the resulting normal and tangential force fields $\bb_n$ and $\bb_t$.
    }
    \label{fig:figure-method-friction}
\end{figure}

In this section, we extend our previous work on a fully Eulerian model for frictionless
contact~\cite{lorez_eulerian_2024} to frictional contact using penalty-based body forces for both
normal and tangential traction components.
We present the methodology for two body contact, however, any number of bodies can be included in a
pairwise manner. Consider two deformable bodies $\mathcal{B}^k$ and $\mathcal{B}^l$ which come into
contact (see \cref{fig:figure-method-friction}a). Following the framework presented in \cref{sec:method}, each body is defined by its own
set of field variables $\phi$ and $\xi$. 
% The non-uniform reference maps $\xi^k$ and $\xi^l$ necessitate a coupling mechanism of the two
% bodies to ensure that the bodies are seperated.

Unlike previous works, particularly in the context of fluid-structure interaction
(FSI)~\cite{mao_3d_2024,rath_efficient_2024,valkov_eulerian_2015}, we employ a multi-velocity field
approach. While a unified velocity field practically eliminates the need for additional contact
forces due to the enforced continuity of the velocity field naturally preventing interpenetration, using
separate velocity fields allows us to explicitly define the contact law. Furthermore, this approach facilitates sliding
between bodies, which is generally not possible with a unified velocity field.
To do so, we introduce a penalty body force field in \cref{eq:equilibrium} to minimize intersection
of the phase-field variables $\phi^k$, $\phi^l$ beyond their half-value $\phi = 0.5$, and hence,
avoid inter-penetration of the bodies.

In the following, we use superscripts for the body index, subscripts $n$ and $\tau$
to indicate normal and frictional components, and a further subscript $t$ indicating the time step. 

\subsection{Normal contact formulation}
\label{sec:normal-contact}
To detect contact we define
a scalar intersection variable for two bodies $\mathcal{B}_k$ and $\mathcal{B}_l$ as
\begin{equation}
	\theta^{kl} := \left\langle \phi^k\cdot\phi^l - \dfrac{1}{4} \right\rangle^+ ~,
	\label{eq:intersection-variable}
\end{equation}
where $\langle \cdot \rangle^+$ is the Macaulay bracket denoting the positive part of the argument. This intersection variable is illustrated for a 1D cut across the interface in normal direction in~\cref{fig:figure-method-friction}b. Due to the symmetry of two phase-fields, the intersection variable
$\theta^{kl}$ is non-zero only in the immediate neighbourhood of regions where $\phi^k>0.5$ and $\phi^l>0.5$. In contrast to
traditional contact formulations defined upon the gap function, the intersection variable is not
directly proportional to the overlapping distance (it has an upper limit of $3/4$). Instead, the total
reaction will be more sensitive to the volume of the contact zone. 

\Cref{eq:intersection-variable} defines a subset of the domain where contact forces are introduced. To
construct a normal and tangential force field $b_n$ and $b_t$, we make use of the available surface
normals being the normalized gradients of the respective phase-field variables:
\begin{equation}
	\boldsymbol{n}^k = \dfrac{\nabla\phi^k}{\vert\nabla\phi^k\vert} ~, \quad \boldsymbol{n}^l = \dfrac{\nabla\phi^l}{\vert\nabla\phi^l\vert} ~.
	\label{eq:normal-definition}
\end{equation}

Introducing a normal penalty constant $\kappa$, the normal force field as proposed in~\cite{lorez_eulerian_2024} is defined as (\cref{fig:figure-method-friction}d)
\begin{equation}
	\boldsymbol{b}_n^k = \kappa \cdot \boldsymbol{n}^l \cdot \theta^{kl} ~.
	\label{eq:normal-force}
\end{equation}

\subsection{Tangential traction and frictional contact}
\label{sec:frictional-contact}

While our proposed method is agnostic to the choice of the friction law, for demonstration purposes, we focus on
Coulomb friction in this work. Therefore we assume the tangential traction is
proportional to the normal traction, with the proportionality constant being the friction
coefficient $\mu$. For a given slip rate $\boldsymbol\delta$ and its magnitude $\norm\delta$, the following relations apply:
\begin{gather}
    \norm{\bb_\tau} \leq \mu \norm{\bb_n} \quad \text{and}\\
    \norm\delta \cdot (\mu \norm{\bb_n} - \norm{\bb_\tau}) = 0 \quad ,
\end{gather}
which define the admissible region in the $\norm\delta$-$\lambda$ space as a Heaviside step function (see dashed black line in~\cref{fig:figure-method-friction}c).

We implement a penalty approach to enforce this condition, defining a tangential penalty constant
$\beta$ and a tangential penalty function $\lambda(\norm\delta, \beta): \mathbb{R}^+ \rightarrow [0, 1]$ for the magnitude
of $\bb_t$. The penalizing function $\lambda$ should be a rapidly increasing continuous function of the slip
rate, approximating the Heaviside step function:
\begin{equation}
	\lambda(\norm\delta, \beta) \sim H(\delta) := \begin{cases}
		1 & \text{if } \norm\delta > 0 \\
        0 & \text{if } \norm\delta = 0\end{cases}
\end{equation}

In this work, we use a simple bilinear function for $\lambda$, ensuring that the penalty increases rapidly with a slope of $\partial\lambda / \partial\delta = \beta$. The function is displayed in~\cref{fig:figure-method-friction}b and reaches a value of $1$ when $\norm\delta \geq 1/\beta$, defined as:
\begin{equation}
    \lambda(\norm\delta, \beta) = \begin{cases}
        \beta \norm\delta & \text{if } \norm\delta < 1/\beta, \\
        1 & \text{otherwise},
    \end{cases}
\end{equation}
or equivalently,
\begin{equation}
    \lambda(\norm\delta,\beta) = \min\left(\beta\norm\delta, 1\right) .
\end{equation}

To implement a frictional force field, we leverage the known relative velocity between the two
bodies that is naturally available due to the multi-velocity field formulation
and the normals obtained from \cref{eq:normal-definition}. The projection of
the relative velocity $\overline{\vv}_{kl} = \vv_k - \vv_l$ onto the tangential plane returns the slip
rate $\boldsymbol{\delta}$:
\begin{equation}
    \boldsymbol{\delta}^{kl} = \mathbf{P} \cdot \overline{\vv}^{kl} ~, \quad \mathbf{P} = \mathbf{I} - \boldsymbol{n}^l \otimes \boldsymbol{n}^l ~,
\end{equation}
where $\mathbf{P}$ is the projection operator onto the tangential plane, and $\mathbf{I}$ is the
identity tensor. Figures~\ref{fig:figure-method-friction}(a~\&~d) provide an overview. The tangential force field acts opposite to the slip rate and is defined as:
\begin{equation}
	\boldsymbol{b}_t^k(\phi^k, \phi^l, \vv^k, \vv^l, \kappa, \beta) = 
    -\lambda(\norm{\boldsymbol\delta^{kl}}, \beta)\cdot \mu\cdot \underbrace{\kappa\cdot \theta^{kl}}_{\norm*{\bb_n^k}}
    \cdot \dfrac{\boldsymbol\delta^{kl}}{\norm{\boldsymbol\delta^{kl}}} ~.
	\label{eq:tangential-force}
\end{equation}

\Cref{eq:tangential-force} expresses the tangential force as a direct function of the velocity fields
$\vv$ and the phase-fields $\phi$. The forces from \cref{eq:normal-force} and
\cref{eq:tangential-force} are incorporated into \cref{eq:equilibrium} to find the field variables
$\phi$ and $\boldsymbol{\xi}$ for both bodies in a fully implicit way. For a multi-body system,
contact force fields are included for all pairs of bodies in contact, and the total force field is the sum
of all pair-wise contact forces. It is crucial to choose the penalty constants $\kappa$ and $\beta$
such that the contact forces are sufficiently large to minimize inter-penetration in the normal
direction and limit sliding for forces not on the Coulomb slip plane. The large gradients in the
energy landscape inevitably lead to an ill-conditioned system, which is challenging to solve. To
address this, we use an approach related to the interior point
method~\cite{houssein_regularized_2023} and the continuation method in topology
optimization~\cite{rojas-labanda_automatic_2015}, which involves dynamically relaxing the constraint, progressively guiding the
solution towards the feasible region. 
However, due to the non-convex nature of the energy landscape there is a possibility that the method
may settle in a suboptimal local minimum rather than the desired one. The details are presented in~\cref{sec:solver-details}.

% ----------------------------------------------------------------------------------------------------------------

\section{Numerical implementation}
\label{sec:numerical_implementation}

\subsection{Discretization}

For a system of $N$ bodies, we construct a monolithic variational problem including the balance
of linear momentum~(\cref{eq:equilibrium}), the phase-field evolution~(\cref{eq:cahn-hilliard}), and
the reference map evolution~(\cref{eq:reference-map-evolution}), for each body $k$.
Any notion of time $t$ refers to a pseudo time.
The system is solved by the finite element method, where the weak form of the
equations is discretized using linear shape functions.

Suppose $\mathcal{S}_{f^k}$ and $\mathcal{V}_{\delta f^k}$ denote the space of trial and test functions for the functions $f^k \in \{\phi^k, \xi^k, q^k\}$ such that:
\begin{align}
    \mathcal{S}_{f^k} &= \left\{ f^k \in H^1(\Omega) \mid f^k = f^k_D \text{ on } \partial\Omega \right\} ~, \quad \\
    \mathcal{V}_{\delta f^k} &= \left\{ \delta f^k \in H^1(\Omega) \mid \delta f^k = 0 \text{ on } \partial\Omega \right\} ~.
\end{align}

The discrete variational form of the linear momentum balance equation at the next time $t+1$ is given by:
\begin{equation}
    \int_\Omega {\phi_{t+1}^k \boldsymbol{\sigma}^k_{t+1}} \colon \nabla\delta\xi^k ~\mathrm{d}\Omega - 
    \int_\Omega \boldsymbol{b}_{n, t+1}^k \cdot \delta\xi^k ~\mathrm{d}\Omega
    - \int_\Omega \boldsymbol{b}_{\tau, t+1}^k \cdot \delta\xi^k ~\mathrm{d}\Omega
     = 0 ~,
     \label{eq:form-linear-momentum}
\end{equation}
where $\colon$ denotes the double contraction of two second-order tensors (inner product). The penalty force densities
$\bb_n^k$ and $\bb_{\tau}$ are: 
\begin{align}
    \bb_{n, t+1}^k &= -\sum_{l\in N} \kappa \cdot \boldsymbol{n}^l(\phi^l_{t+1}) \cdot \theta^{kl}(\phi^k_{t+1}, \phi^l_{t+1})~, \\
    \bb_{\tau, t+1}^k &= -\sum_{l\in N} \lambda(\norm{\boldsymbol{\delta}^{kl}}, \beta)\cdot \mu\cdot \kappa\cdot \theta^{kl}(\phi^k_{t+1}, \phi^l_{t+1})\cdot \dfrac{\boldsymbol\delta^{kl}}{||\boldsymbol\delta^{kl}||} ~.
\end{align}

\begin{remark}
    We consider static equilibrium at $t+1$. The tangential force density $\bb_t$ contains the slip
    rate $\boldsymbol\delta$ which is a function of the material velocities. For this, we use the
    intra-step velocity between $t \rightarrow t+1$, therefore assuming a constant velocity in the
    timestep. This approach necessitates sufficiently small timesteps. A more accurate approximation of
    the velocity at $t+1$ would be $(\vv_{t+1/2} + \vv_{t+3/2})/2$ which is unavailable in the
    currently employed scheme.
\end{remark}

For temporal discretization of \cref{eq:cahn-hilliard,eq:reference-map-evolution}, we use the
Crank-Nicolson method. 
The semi-discrete form of the reference map evolution yields the velocity field $\vv_{t+1/2} \in H^0(\Omega)$ as a function of the reference map $(\xi_{t+1}, \xi_t) \in (H^1(\Omega))^2$:
\begin{equation}
    \vv_{t+1/2}^k = -2 \left[\nabla\left(\frac{\xi^k_{t+1}+\xi^k_t}{2}\right)\right]^{-1} \dfrac{\xi_{t+1}^k - \xi_t^k}{\Delta t} ~.
\end{equation}

To solve the fourth-order Cahn-Hilliard equation with first-order Finite Elements, we cast \cref{eq:cahn-hilliard} into a system of two second-order equations, introducing the auxiliary variable $q = \delta\mathcal{E}_{\text{PF}} / \delta\phi$~\cite{lorez_eulerian_2024}\footnote{In our previous work~\cite{lorez_eulerian_2024}, the auxiliary function was denoted with the symbol $\mu$. Here, we changed it to not confuse it with the friction coefficient.}:
\begin{gather}
    \int_\Omega \dfrac{\phi_{t+1} - \phi_t}{\Delta t} \delta\phi ~\mathrm{d}\Omega + \int_\Omega \vv_{t+1/2} \cdot \nabla\left(\frac{\phi_{t+1}+\phi_{t}}{2}\right) \delta\phi ~\mathrm{d}\Omega
    + \mathcal{M} \int_\Omega \left(\nabla q\cdot\nabla\delta\phi\right) ~\mathrm{d}\Omega = 0 ~,
    \label{eq:form-phase-field} \\
    \int_\Omega  q \delta q ~\mathrm{d}\Omega - \int_\Omega \epsilon^2 ~\nabla\left(\frac{\phi_{t+1} + \phi_t}{2}\right) \cdot \nabla\delta q ~\mathrm{d}\Omega - \int_\Omega \left(\frac{g'(\phi_{t+1}) + g'(\phi_t)}{2}\right)\delta q ~\mathrm{d}\Omega = 0 ~.
    \label{eq:form-chemical-energy}
\end{gather}

Then, the problem reads: Find the functions $(\xi_{t+1}^k, \phi_{t+1}^k,  q_{t+1}^k) \in
\mathcal{V} = (\mathcal{H}^1(\Omega))^3$ for every body such that
\cref{eq:form-linear-momentum,eq:form-phase-field,eq:form-chemical-energy} are satisfied. We implement the monolithic fully
coupled system using FEniCS~\cite{logg_automated_2012}.
Schematically, for a single body, the system to solve looks as follows:
\begin{equation}
	\underbrace{
		\begin{bmatrix}
            \mathbf{K_\xi}    & \mathbf{K_{\xi\phi}}                  & 0                    \\
			                  & \mathrm{K_\phi}    & \mathrm{K_{\phi q}} \\
			                  &                    & \mathrm{K_q}
		\end{bmatrix}
	}_{\mathbf{K^{(k)}}}
	\cdot
	\underbrace{
		\begin{bmatrix}
			\vector{\xi} \\ \phi \\ q
		\end{bmatrix}
	}_{\mathbf{u^{(k)}}}
	- \underbrace{\begin{bmatrix}
			\vector{b} \\ 0 \\ 0
		\end{bmatrix}}_\mathbf{f^{(k)}}
	= \vector{0}
\end{equation}

Considering multiple bodies, the system is constructed as follows:
\begin{equation}
	\begin{bmatrix}
        [\mathbf{K^{(k)}}] & 0 \\  & [\mathbf{K^{(l)}}]
	\end{bmatrix}
	\cdot
	\begin{bmatrix}
		\mathbf{u^{(k)}} \\ \mathbf{u^{(l)}}
	\end{bmatrix}
	- \begin{bmatrix}
		\mathbf{f^{(k)}} \\
		\mathbf{f^{(l)}}
	\end{bmatrix}
	= \vector{0}
\end{equation}

\subsection{Solver details}
\label{sec:solver-details}

The resulting non-linear system is solved using PETSc's \textit{SNES} solver, employing the Newton method with
line search, \textit{MUMPS} for the linearized subproblems, and \textit{AMG} pre-conditioning~\cite{balay_petsctao_2023}.

To resolve the frictional inequality to high accuracy, in particular to retrieve the slip-stick
differentiation, relatively large penalty coefficients $\beta$ are required. However, for large penalty
coefficients $\beta$, the frictional penalty forces in the contact formulation lead to a highly
ill-conditioned system, which is challenging to solve. 
Furthermore, the problem is magnified by the reduced admissible solution space constrained by the
Lorentz cone.

Therefore, to reliably obtain a solution, we decompose the problem into two subproblems: First, we address
the stick problem, where the frictional force is linearly proportional to the slip rate. Second, we
progressively allow slip to return to the Coulomb slip surface, which is effectively equivalent to a
return mapping algorithm. 

\begin{algorithm}
    \caption{Continuation method and return mapping}
    \label{alg:continuation-method}
    \KwIn{Non-linear minimization problem $\mathscr{P}$, starting point $\mathbf{u}_0$, target penalty parameter $\beta_t$, increment factor $\gamma^+_\beta$, reduction
    factor $\gamma^-_\beta$, slip mixing increment $\Delta\alpha$.}
    \KwOut{Solution $\mathbf{u}$}
    $\beta \gets \beta_t$ \\
    $\alpha \gets 0$ \\
    $\mathbf{u} \gets \mathbf{u}_0$ \\
    \While{}{
        Backup $\mathbf{u}$: $\mathbf{u_*} \gets \mathbf{u}$ \\
        \uIf{
            Minimize $\mathscr{P}$: $\mathbf{u} \gets \min\mathscr{P}(\mathbf{u}, \beta)$
        }{
            \If{$\beta \geq \beta_t$}{
                {\bf break}
            }
            Increase penalty $\beta \gets \gamma^+_\beta\beta$ \\
        }
        \Else{
            Revert $\mathbf{u} \gets \mathbf{u_*}$ \\
            Reduce penalty $\beta \gets \gamma^-_\beta\beta$
        }
    }
    \While{$\alpha < 1$}{
        Minimize $\mathscr{P}$: $\mathbf{u} \gets \min\mathscr{P}(\mathbf{u}, \alpha)$ \\
        Increase slip mixing $\alpha \gets \alpha + \Delta\alpha$
    }
    \Return $\mathbf{u}$
\end{algorithm}

\subsubsection{Continuation method}

To solve the stick-only problem, we employ the continuation method, which incrementally increases
the constraints to guide the solution toward the feasible region, if necessary. This approach helps
mitigate ill-conditioning by avoiding large, abrupt changes in the solution space. The continuation
method can be viewed as a generalization of the penalty method, where the penalty parameter is
dynamically adjusted to approach the solution of the original
problem~\cite{rojas-labanda_automatic_2015}. It can also be seen as a simplification of the
augmented Lagrangian method~\cite{belytschko_nonlinear_2014}, and shares principles with the
interior point method (IPM).

However, due to the non-convex nature of the frictional contact problem, the continuation method may
not always guarantee a unique solution. While the method is generally effective at finding the
global optimum, our primary interest lies in identifying the local minimum, particularly by
leveraging the variation of the previous solution. Therefore, caution is warranted when applying the
continuation method in this context, as there is a risk of converging to an unintended local
minimum. To reduce this risk, we start with the target penalty parameter $\beta_t$ and decrease it
only if the solver fails to converge. The procedure is outlined in Algorithm~\ref{alg:continuation-method}.

\subsubsection{Return mapping}

After solving the stick problem, we introduce slip by incrementally increasing the slip mixing
parameter $\alpha$ from 0 to 1. This scalar parameter linearly interpolates between the stick and
slip solutions, allowing for a smooth transition between these two states:
\begin{gather}
    \lambda = (1-\alpha) \lambda_{\text{stick}} + \alpha \lambda_{\text{Coulomb}} ~, \\
    \lambda_{\text{stick}}(\delta, \beta) = \beta\delta ~, \quad \lambda_{\text{Coulomb}}(\delta, \beta) = \min\left(\beta\delta, 1\right) ~.
\end{gather}

The return mapping process is demonstrated in \cref{sec:return-mapping-illustration}, and the method
is outlined in Algorithm~\ref{alg:continuation-method}. We acknowledge that the current approach may
not be optimal, and potential improvements are identified as areas for future work.

\section{Examples}
\label{sec:examples}

In this section, we present three numerical examples to validate and showcase the proposed
methodology. Throughout the examples, we use dimensionless units for space and time.
We start by validating the proposed methodology for the Cattaneo-Mindlin problem, a classical
benchmark problem for frictional contact for which an analytical solution is available~\cite{mindlin_compliance_1949,johnson_contact_1987}. We first
demonstrate that our method yields the correct traction profile, and investigate the return mapping process
to illustrate the internal solution process. We also show the path dependency by
performing a full cycle of forward and backward sliding, and show that a reasonable hysteresis loop is obtained
when sliding the bodies back and forth. Next, we simulate the ironing problem to show large
sliding~\cite{cavalieri_numerical_2015,puso_mortar_2004,de_lorenzis_large_2011}. 
In the last example, we analyze a system with two deformable solids and multiple contact patches.
For all shown examples, we use dimensionless units.

\subsection{Cattaneo-Mindlin problem}
\label{sec:cattaneo-mindlin-problem}

\subsubsection{Tangential traction profile}
\label{sec:cattmind:traction-profile}

\begin{figure}
	\begin{center}
		\includegraphics{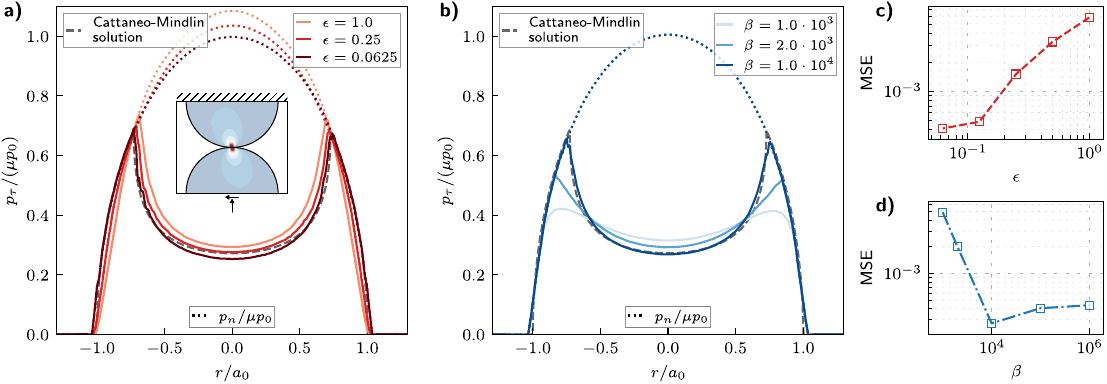}
	\end{center}
	\caption{Analysis of the tangential traction profile for different values of diffuse interface
		width $\epsilon$ and frictional penalty coefficients $\beta$ in the Cattaneo-Mindlin
		problem. For all displayed cases, the normal penalty is $\kappa = 200$, and the mobility
		parameter is $\mathcal{M} = 10^{-6}$. \textbf{(a)} Normalized traction profile for varying
		$\epsilon$. The analytical solution is shown as the dashed black line. The dotted lines are
		the normal traction ($\mu p_n$). \textbf{(b)} Normalized traction profile for varying
		$\beta$ (for $\epsilon = 0.125$). \textbf{(c, d)} Mean squared error (MSE) between the
		resulting traction profile and the analytical solution for increasing $\epsilon$ and
		$\beta$, respectively.}
	\label{fig:cattaneo-mindlin-problem}
\end{figure}

The Cattaneo-Mindlin problem is a classical benchmark problem for frictional
contact~\cite{mindlin_compliance_1949,johnson_contact_1987}. The problem consists of two linear
elastic cylinders in contact, pushed together and then sheared. The setup is shown in the inset of
\cref{fig:cattaneo-mindlin-problem}a. Equivalent to the Hertzian solution,
the analytical solution for the tangential traction profile, given the total normal and tangential
reaction force, is known. We use this problem to test the proposed methodology for frictional
contact with partial slip, specifically focusing on whether the differentiation between sliding and sticking regions is correctly captured. 
The two cylinders are modeled as linear elastic, $i.e.$
\begin{equation}
    \boldsymbol{\sigma} = \lambda \mathrm{tr}(\mathbf{\boldsymbol{\varepsilon}}) \mathbf{I} + 2\mu \mathbf{\boldsymbol{\varepsilon}} ~,
    \quad \boldsymbol{\varepsilon} = \dfrac{1}{2} \left(\mathbf{F} + \mathbf{F}^T\right) - \mathbf{I} ~.
    \label{eq:linear-elastic-stress}
\end{equation}
They have a radius of $R_0 = R_1 = 10$, Young's modulus $E_0 = E_1 = 0.2$, and Poisson's ratio
$\nu_0 = \nu_1 = 0.2$. The friction coefficient is $\mu = 0.5$. We clamp the top cylinder at its
top boundary and apply to the bottom cylinder first, a vertical displacement $\bar{u}_y = 0.03$ and then a
horizontal displacement of $\bar{u}_x = 0.01$.

The analytical solution of the traction profile is characterized by the effective young's modulus $E^*$, the contact zone width $a$,
the stick-slip transition width $c$, the normal and tangential reaction forces $P_n$ and $P_\tau$, and
the friction coefficient $\mu$:

\begin{gather}
    E^* = \left(\dfrac{1-\nu_0^2}{E_0} + \dfrac{1-\nu_1^2}{E_1}\right)^{-1} ~, \quad R = \left(\dfrac{1}{R_0} + \dfrac{1}{R_1}\right)^{-1}~, \quad
    a = \left( \frac{4 P_n R}{\pi E^*}  \right)^{1/2} ~, \quad c = a \left( 1 - \frac{P_\tau}{\mu P_n} \right)^{1/2}
\end{gather}

Then, the analytical solution for the normal and tangential traction profiles were found to be:
\begin{align}
    p_n(r) &= \dfrac{2 P_n}{\pi a^2} \sqrt{a^2 - r^2}~, \\
    p_\tau(r) &= \frac{2 \mu P_n}{\pi a^2} \left(\sqrt{a^2 - r^2} - H(c^2-r^2)\sqrt{c^2 - r^2}\right)~. 
\end{align}

Figure \ref{fig:cattaneo-mindlin-problem} demonstrates the strong agreement between our numerical
experiments and the analytical solution. To achieve this, we define the traction as the integrated
force density perpendicular to the surface, denoted as $p_\tau = \int_{n} \bb_\tau ~\mathrm{d}n$, and the
reaction force as the volume integral $P_\tau = \int_\Omega \bb_\tau ~\mathrm{d}\Omega$. We normalize the
traction profile using the maximum Hertzian traction at $r=0$. For a review of the normal traction
profile, the reader is referred to~\cite{lorez_eulerian_2024}.

In \cref{fig:cattaneo-mindlin-problem}a, we display the traction for various values of the diffuse
interface width $\epsilon$, with the analytical solution represented by the dashed black line. The
normal traction (scaled with $\mu$) is depicted as the dotted line. As $\epsilon$ decreases, the
numerical solution approaches the analytical solution. It is important to note that the
normalization obscures the fact that the absolute traction is lower for large $\epsilon$ than the smaller $\epsilon$ due to
the softer implied contact stiffness (for equal $\kappa$).
In \cref{fig:cattaneo-mindlin-problem}b, we show the traction for varying frictional penalty 
coefficients $\beta$ for a fixed $\epsilon = 0.125$. It is evident that the solution is converging
for sufficiently large $\beta$.
The mean squared error (MSE) between the numerical and analytical solutions is shown in
\cref{fig:cattaneo-mindlin-problem}(c, d). The error does not decrease further for $\beta > 10^4$
due to limitations from the discretization. Further $h$-refinement would be required to reduce the
error further.
Overall, the results demonstrate that the proposed methodology is capable to accurately capture the
tangential traction profile for the Cattaneo-Mindlin problem.

\subsubsection{Return mapping illustration}
\label{sec:return-mapping-illustration}

\begin{figure}  % figure alpha illustration
	\begin{center}
		\includegraphics{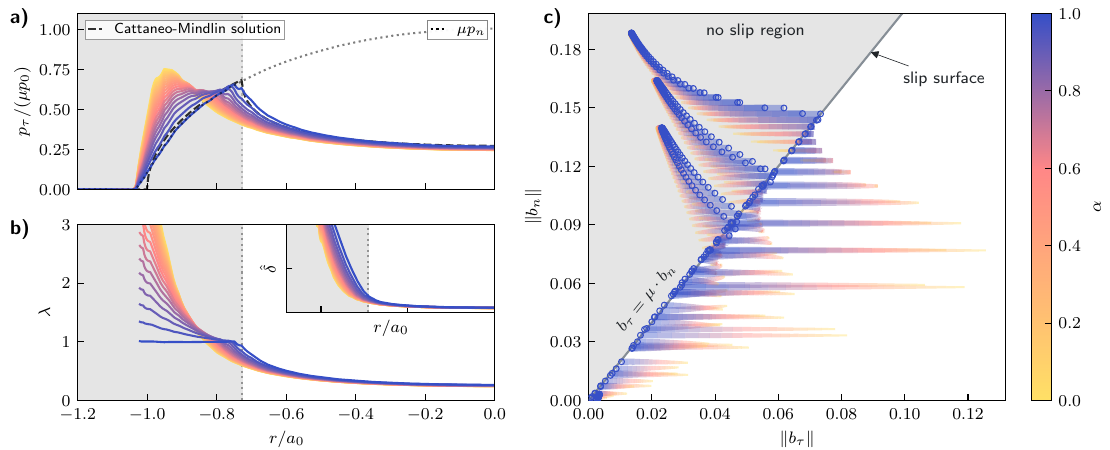}
	\end{center}
	\caption{Illustration of the implicit return mapping on the example of a single Cattaneo-Mindlin
        simulation with $\epsilon = 0.125$ and $\beta = 10^4$. \textbf{(a)} Intra-step evolution of
        the traction profile with the implicit return mapping. Starting from a strictly sticking
        interface ($\alpha=0$) to a partially sliding interface ($\alpha=1$). The dashed black line
        is the analytical solution. \textbf{(b)} The local effective slip rate $\hat\delta$ for every
        position in the contact patch. \textbf{(c)} The local absolute contact force densities $b_n$
        and $b_\tau$ for every node in the contact zone, from $\alpha=0$ to $\alpha=1$. Every trace
        represents one node, and the round markers indicate the final state. For nodes where the tangential force initially exceeds the valid value
        of $\mu b_n$, the tangential component is effectively reduced to the slip surface.
	}
	\label{fig:return-mapping}
\end{figure}

\Cref{fig:return-mapping} illustrates the implicit return mapping for a single Cattaneo-Mindlin
simulation with $\epsilon = 0.125$ and $\beta = 10^4$. The figure shows the intra-step evolution from $\alpha=0$ (yellow) to
$\alpha=1$ (blue) of
the traction profile (\cref{fig:return-mapping}a), the value of $\lambda$, and the local slip rate
$\delta$ in the contact zone (\cref{fig:return-mapping}b), and the local absolute force densities
$b_n$ and $b_\tau$ for every node in the contact zone (\cref{fig:return-mapping}c).
The figure demonstrates the correct differentiation between sticking and sliding
regions, as well as the correct enforcement of the frictional law.
In \cref{fig:return-mapping}c, for nodes where the tangential force initially exceeds the admissible
maximum of $\mu b_n$, the tangential component is effectively reduced to the slip surface (Lorentz cone).

In the left part of \cref{fig:return-mapping}, the tractions, being the integrated force densities across the contact patch, and the frictional function $\lambda(r)$ are computed as:
\begin{equation}
    p_n(r) = \int_y b_n(r, y) ~\mathrm{d}y ~, 
    \quad p_\tau(r) = \int_y b_\tau(r, y) ~\mathrm{d}y ~, 
    \quad \lambda(r) = \frac{p_\tau(r)}{\mu p_n(r)} ~,
    \label{eq:traction-across-contact-patch}
\end{equation}
and the effective slip rate $\hat\delta$ according to \cref{eq:effective-slip-rate}.
For $\alpha=0$, considered to be an artificial state of full stick, the tangential traction is
maximal at the edge of the contact area. 
When the Coulomb friction law is consecutively mixed in by increasing $\alpha$, the tangential
traction profile transitions to the analytical solution. 

The right part of \cref{fig:return-mapping} illustrates the local force densities $b_n$ and $b_\tau$
for every node in the contact zone. The three distinct clusters of nodes represent three layers of
nodes in normal direction inside the volumetric contact patch. While for $\alpha=0$, the points span
the entire space, for $\alpha=1$, all points are inside the admissible friction cone or lay exactly
on the slip surface $b_n = \mu b_\tau$. Further we observe that most points shift in horizontal
direction, meaning that the return mapping process works mainly by implicitly altering the
tangential force density by altering the horizontal component of displacement, while maintaining a
constant normal pressure.

\subsubsection{Hysteresis loop}
\label{sec:hysteresis-loop}

\begin{figure}
    \begin{center}
        \includegraphics{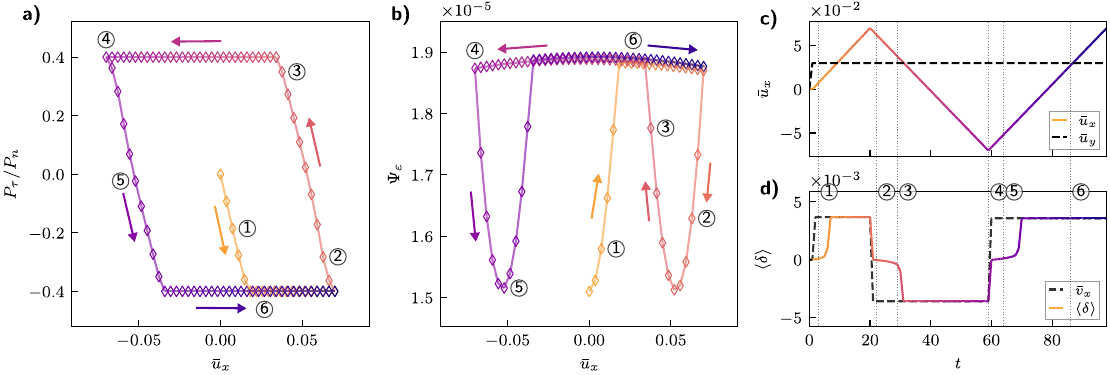}
    \end{center}
    \caption{Hysteresis loop for the Cattaneo-Mindlin problem. The figure shows the
        tangential reaction force $P_\tau$ for the forward and backward loading paths. The
        hysteresis loop is a result of the path-dependent energy dissipation in the frictional
        contact problem. \circled{1} Initial tangential loading; partial slip. \circled{2} Point of
        reversal; stick and unloading. \circled{3} On the brink of slipping. \circled{4} Full slip.
        \circled{5} Neutral total friction force; partial slip. \circled{6} Full slip.
        \textbf{(a)} Tangential reaction force in time. The $x$-axis shows the applied position at
        the bottom of the bottom cylinder, while the $y$-axis shows the normalized reaction force
        $P_\tau / P_n$.
        \textbf{(b)} Integrated strain energy.
        \textbf{(c)} Applied motion at the bottom edge of the bottom cylinder. First, a vertical
        velocity is applied, followed by a full cycle of forward and backward horizontal motion.
        \textbf{(d)} Averaged slip rate at the contact interface, in comparison to the applied
        velocity of the bottom edge of the bottom cylinder.
    }
    \label{fig:cattmind-hysteresis}
\end{figure}

To illustrate path-dependent energy dissipation in the frictional contact problem, we consider the
same setup as in \cref{sec:cattmind:traction-profile} and conduct a complete cycle of forward and
backward shearing motion. The used parameters are $E=0.2$, $\nu=0.2$, $\mu=0.4$, $\epsilon=0.125$, $\mathcal{M}=10^{-8}$,
$\kappa=300$, and $\beta=4\cdot 10^4$. We prescribe the motion of the bottom half cylinder, which is
depicted in \cref{fig:cattmind-hysteresis}c. Initially, the bottom cylinder is moved vertically by
$\bar{u}_y = 0.03$, followed by a horizontal motion within $\bar{u}_x \in [-0.07, 0.07]$. Throughout
\cref{fig:cattmind-hysteresis}, we indicate six characteristic stages with annotations $1$ to $6$.

\Cref{fig:cattmind-hysteresis}(a,b) show the hysteresis loop for the relative tangential reaction
force $P_\tau/P_n$ and the strain energy $\Psi_{\varepsilon}$ respectively.
The hysteresis loop reveals an expected
path-dependent energy dissipation, where the forward and backward loading paths are clearly
distinguishable.
In addition, \cref{fig:cattmind-hysteresis}d shows the average slip rate $\langle \delta
\rangle$ in the contact patch in comparison to the applied bottom boundary velocity $\bar{v}_x$. 
To compute the average slip rate across the contact patch, we first define an effective slip rate
across its thickness using a weighted average of the slip rate $\delta(r, y)$ and the tangential
force density $b_\tau(r, y)$;
\begin{equation}
    \hat\delta(r) = \dfrac{\int_y \delta(r, y) \cdot b_\tau(r, y) ~\mathrm{d}y}{\int_y b_\tau(r, y) ~\mathrm{d}y} \quad ,
    \label{eq:effective-slip-rate}
\end{equation}
and then average the effective slip rate across the width of the contact patch:
\begin{equation}
    \langle\delta\rangle = \dfrac{\int_r \hat\delta(r) ~\mathrm{d}r}{\int_r 1 ~\mathrm{d}r} \quad .
    \label{eq:average-slip-rate}
\end{equation}

The discontinuities in $\langle \delta \rangle$ align with the transition points in the applied
velocity, showing characteristic jumps near stage $2$ and $4$ when the motion
is reversed. Upon the reversal of the external motion, $\langle\delta\rangle$ stays close to $0$,
meaning a sticking interface; then it grows in magnitude until the full contact patch is in a
sliding state and the average slip rate equals the externally applied velocity. This behavior
highlights a stick-slip phenomenon at the interface, where the contact patch transitions from full
stick to full slip.

\subsection{Ironing problem}
\label{sec:ironing-problem}

\begin{figure}
    \begin{center}
        \includegraphics{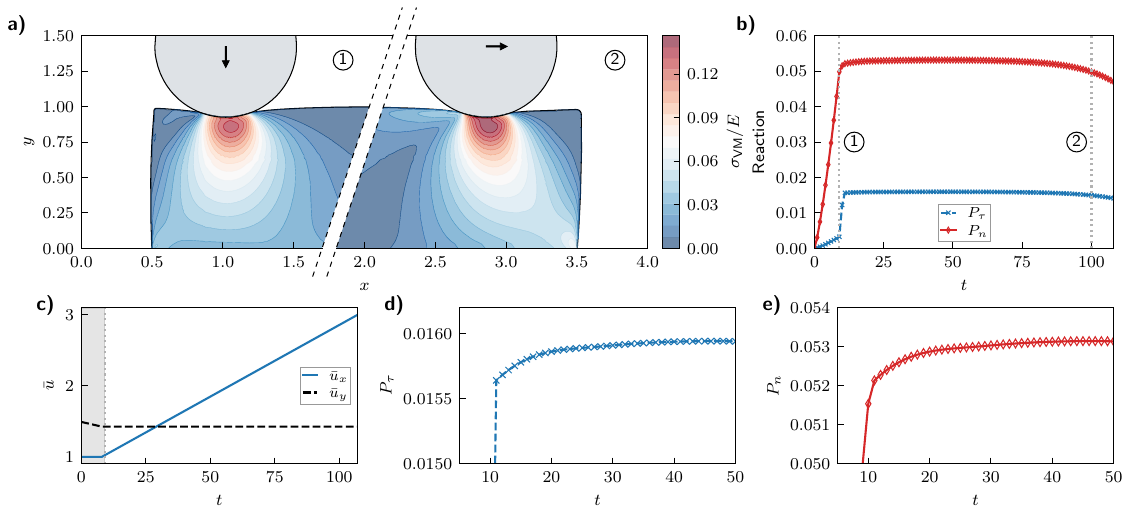}
    \end{center}
    \caption{Ironing problem.
        \textbf{(a)} Problem setup and von Mises stress intensity for two
        time instances: \circled{1} shows the state when the die has reached its final vertical position, and \circled{2} depicts the die at a later time during horizontal sliding. 
        \textbf{(b)} Reaction forces $P_\tau$ (tangential) and $P_n$ (normal) over the course of the simulation. 
        \textbf{(c)} Prescribed motion of the rigid die. First a vertical displacement
        of $\bar{u}_y=-0.075$ is applied in 10 steps, followed by a horizontal displacement of $\bar{u}_x=2.0$ in a further 100
        steps.
        \textbf{(d, e)} Zoomed-in view of the reaction forces $P_\tau$ and $P_n$.
    }
    \label{fig:ironing-result}
\end{figure}

The ironing problem is a benchmark for testing large sliding in contact formulations~\cite{cavalieri_numerical_2015,puso_mortar_2004,de_lorenzis_large_2011}. In
this setup, a rigid cylindrical die with a radius of $R=0.5$ is pressed into a deformable box with
dimensions $3 \times 1$, which is fixed at the bottom (see \cref{fig:ironing-result}). The box material is modeled as a Neo-Hookean
hyperelastic material law with Young's modulus $E=1.0$ and Poisson's ratio $\nu=0.3$.
Its strain energy density is given by:
\begin{equation}
    W = \dfrac{\mu}{2} \left(\mathrm{tr}(\mathbf{C}) - d\right) - \mu\ln J + \dfrac{\lambda}{2} (\ln J)^2 ~,
    \label{eq:neo-hookean-engery}
\end{equation}
where $\mu$ and $\lambda$ are the Lamé parameters, $d$ is the dimension of the space,
$J=\mathrm{det}(\mathbf{F})$, and $\mathbf{C} = \mathbf{F}^T\mathbf{F}$ is the right Cauchy-Green
deformation tensor. The Cauchy stress is given by:
\begin{equation}
    \boldsymbol{\sigma} = \dfrac{\mu}{J} \left(\mathbf{F}\mathbf{F^T} - \mathbf{I}\right) +
    \dfrac{\lambda}{J} (\ln J) \mathbf{I} ~.
    \label{eq:cauchy-stress}
\end{equation}
The phase-field model parameters are set to $\epsilon=0.02$ and
$\mathcal{M}=10^{-6}$. The contact is modeled as frictional, with a coefficient of friction
$\mu=0.3$, while the penalty parameters are set to $\kappa=10^3$ and $\beta=10^3$. The die follows a
prescribed motion as depicted in~\cref{fig:ironing-result}c. Initially, a uniform downward
displacement of $\bar{u}_y = -0.075$ is applied over 10 time steps, followed by a horizontal
displacement of $\bar{u}_x = 2.0$ over the subsequent 100 time steps.

We employ the proposed methodology to solve the problem, demonstrating its ability to handle
arbitrarily large sliding. \Cref{fig:ironing-result}a presents the problem setup along with the
von Mises stress distribution. Two instances are shown:
first (left side), after completing the vertical displacement (marked as \circled{1}), and later, towards the end of the simulation (marked as \circled{2}). As expected, the sliding generates horizontal forces, resulting in a
characteristic skewed stress profile. 

In \cref{fig:ironing-result}b, we plot the reaction forces $P_\tau$
and $P_n$ over the course of the simulation. Zoomed-in views are provided in \cref{fig:ironing-result}d
and (e) to enhance clarity along the $y$-axis. 
Initially, we observe an approximately linear increase in the normal reaction up to \circled{1}, accompanied by a small tangential reaction due to the asymmetric box loading. Once the die begins horizontal motion, the frictional reaction quickly saturates at $P_\tau = \mu P_n$. 
Shortly after \circled{1}, a slight increase in the reaction force is observed, attributed to rotational effects induced by the frictional force, which cause greater compression in the system.

While comparable studies employing Lagrangian approaches report small oscillations in reaction forces due to sliding across element boundaries, along with strategies to mitigate these effects~\cite{de_lorenzis_large_2011}, the reactions in our results are perfectly smooth. This is no surprise, since the fixed mesh inherent to an Eulerian approach eliminates the presumed source of these oscillations.
Compared to the results in~\cite{de_lorenzis_large_2011}, the reaction forces in our study are approximately 10\% lower in magnitude. This discrepancy likely arises from the diffuse interface approximation and the deformation of the
softer transition zone ($\phi=0.5\rightarrow1$) at the interface, which acts as a deformation buffer. 
The effect is further pronounced due to the displacement-controlled loading in this setup, as the reaction forces are effectively reduced by the deformation of the \emph{softer} interface. 

Lastly, we note that no convergence issues were encountered with the parameters used in this simulation.

\subsection{Multi-patch frictional contact}
\label{sec:multi-patch-contact-example}

\begin{figure}
    \centering
    \includegraphics{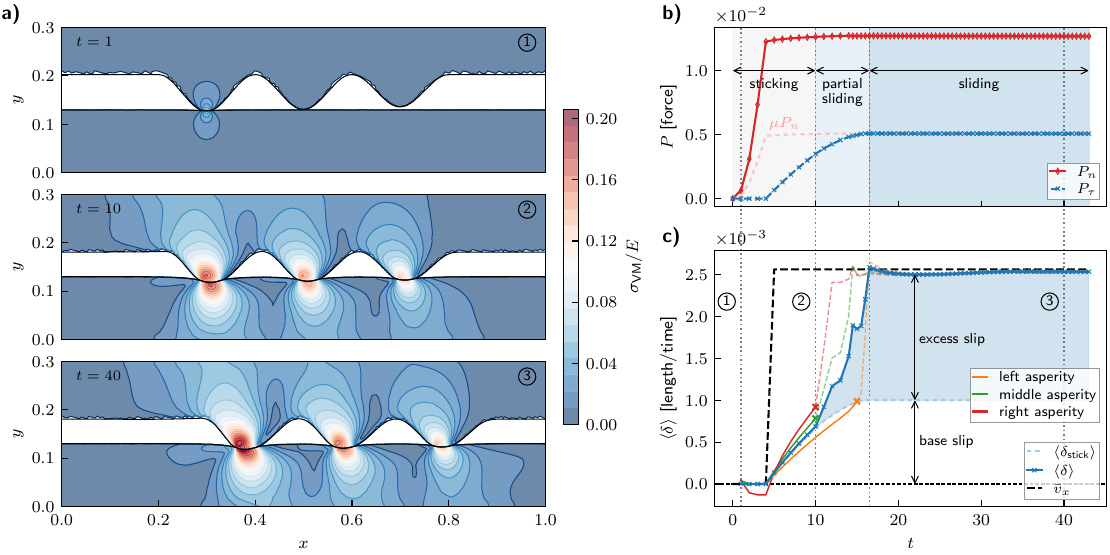}
    \caption{Frictional contact between a body with sinusoidal asperities and a flat plane. 
    \textbf{(a)} Problem setup and von Mises stress intensity for three instances in time.
    \textbf{(b)} Total normal and tangential reaction forces over time.
    \textbf{(c)} Slip rates as a function of time, averaged over the total contact area (blue) and over the contact area of the individual asperities (orange, green, red). The base slip rate is the slip required to reach the sliding surface of Coulomb's law and thus, the erroneous sliding of a sticking interface introduced by the taken penalty approach. For each asperity, the cross marks the moment, when the recorded slip rate exceeds the base slip rate, marking the onset of sliding. The dashed black line shows the prescribed velocity of the top body.
    }
    \label{fig:result-sinusoidal}
\end{figure}

We use the final example to demonstrate the applicability of the proposed
methodology to systems with multiple contact patches between deformable bodies
(\cref{fig:result-sinusoidal}a). The top body features three sinusoidal
asperities of varying magnitudes that interact with the flat surface of the
bottom body.
Initially, the top body is displaced downward by a total vertical displacement of
$0.03$, compressing both bodies. Subsequently, it is moved laterally to the
right with a constant boundary velocity of $\bar{v}_x=2.5\cdot 10^{-3}$,
generating frictional forces at the interface. 

All simulation units are dimensionless. The material is modeled as Neo-Hookean
elastic (see \cref{eq:neo-hookean-engery,eq:cauchy-stress}) with $E=0.6$ and $\nu=0.3$. The phase-field parameters are
$\mathcal{M}=5\cdot 10^{-7}$ and $\epsilon=0.01$, while the contact parameters
are $\kappa=5\cdot 10^3$, $\beta=1\cdot 10^3$, and $\mu=0.4$.

The system setup can be inferred from \cref{fig:result-sinusoidal}. The system
of equations from
~\cref{eq:form-linear-momentum,eq:form-phase-field,eq:form-chemical-energy} is
solved subject to the following boundary conditions:
\begin{itemize}
    \item For the upper body, the reference map $\xi$ is prescribed on the upper boundary as $\xi_{t+1} = \xi_t - \bar{\vv} \Delta t$, corresponding to the imposed boundary motion $\bar{\vv}$. On the lateral sides, only the $x$-component of $\xi$ is prescribed.
    \item For the bottom body, the reference map $\xi$ is fixed at its initial value, $\xi_{t+1} = \xi_t$ at all edges.
\end{itemize}

\Cref{fig:result-sinusoidal}a shows the von Mises stress distribution inside
the body -- where $\phi>0.5$ -- at three time instances. Due to the asymmetry of
the asperities, the stress distribution varies among them. Furthermore,
friction causes a skewed stress distribution.

\Cref{fig:result-sinusoidal}b illustrates the total tangential and normal
reaction forces over time. The tangential force increases continuously as the
top body moves right. Around $t=17$, the tangential force plateaus, indicating that
the interface is in full sliding.

The slip rate over time is shown in \cref{fig:result-sinusoidal}c, where the
average slip rate across the contact area (blue) and individual asperities
(orange, green, red) are plotted. Due to the penalty approach used to enforce 
Coulomb's law, a base slip exists even for a \emph{sticking} interface. This
base slip corresponds to the minimal required slip rate to achieve the observed
frictional force and is determined by the ratio of the average tangential force
density $p_\tau$ to the normal force density $p_n$, scaled inversely by
$\beta\mu$:
\begin{equation}
    p_\tau = \underbrace{\beta\delta}_\lambda (\mu p_n)
    \longrightarrow \langle\delta_{\text{stick}}\rangle = \frac{\langle p_\tau\rangle}{\beta\mu\langle p_n \rangle} \quad ,
\end{equation}
where $\langle\cdot\rangle$ denotes an average over the contact area according
to~\cref{eq:effective-slip-rate,eq:average-slip-rate}.
We consider that sliding begins when the recorded slip exceeds this base slip.
The onset of sliding of the individual asperities is marked by the crosses in
\cref{fig:result-sinusoidal}c.

Sliding initiates at $t=10$ (time instance 2) when the middle and right
asperities begin to slide, while the leftmost and most loaded asperity remains
sticking. After this, the system is in a state of partial slip until
$t\approx17$, at which point the total slip rate matches the top body's velocity
(dashed black line), indicating full sliding. The third time instance shows the
interface in a state of full sliding.
We note that the proposed methodology is capable of effectively capturing interfacial forces at multiple contact patches.

\section{Discussion}
\label{sec:discussion}

In this work, we introduced a novel Eulerian phase-field framework to model frictional contact
between deformable bodies. To the best of our knowledge, this is the first fully functional method that
incorporates frictional contact within a Eulerian phase-field context. By leveraging the reference
map technique in combination with the phase-field method, we developed an approach that captures
large deformations, complex contact interactions, and frictional effects within a Eulerian setting.

A major benefit of the original Eulerian approach~\cite{lorez_eulerian_2024} has been its
simplicity. Contact interactions are implicitly captured through the overlap of diffuse
phase-fields, while penalty-based body forces enforce contact constraints. Extending this method to
handle frictional contact maintains this simplicity, utilizing the available field variables on the
Eulerian mesh to construct frictional penalty forces. 
Lagrangian contact formulations often rely on node-to-node or node-to-segment contact definitions,
which can restrict sliding to small incremental changes and induce subtle oscillations originating
from the discretization of the surfaces in contact~\cite{de_lorenzis_large_2011}. 
The implicit representation of bodies through phase-field variables allows for sliding without
restrictions, enabling the simulation of arbitrarily large sliding motions and the handling of
complex shapes.
Furthermore, the suggested methodology retains all the inherent advantages
of an Eulerian formulation, which is particularly suited for problems involving large deformations,
multiphysical problems, where contact interactions are coupled with other physical or chemical
phenomena, for which the Eulerian description is more natural. 
For example, Eulerian formulations have been prominently used to model fluid-structure
interactions~\cite{frei_eulerian_2016,dunne_eulerian_2006,richter_fully_2013,valkov_eulerian_2015,mokbel_phase-field_2018,mao_interface_2023},
to model growth processes~\cite{li_phase-field_2023,naghibzadeh_surface_2021,jokisaari_phase_2018}
or problems related to topological changes such as merging and separation of bodies~\cite{sass_accurate_2023,benson_contact_2004}.

While we have demonstrated the method using Coulomb friction, there is nothing conceptually limiting
the implementation of more refined friction laws by adapting the frictional penalty force
function~\cref{eq:tangential-force}, such as rate-and-state friction
models~\cite{rice_rate_2001,kammer_linear_2014}. 
As illustrated in~\cref{fig:return-mapping}, the method respects the transition from sticking to sliding,
accurately representing history dependence and energy dissipation associated with frictional
interactions, and the simulations exhibit oscillation-free frictional responses during sliding.

However, several challenges remain or were discovered during the construction of this paper. 
Whereas the proposed method could be used to solve the Coulomb friction problem by a simple
minimization using the Newton method,
accurately capturing the stick-slip differentiation requires a large frictional penalty coefficient $\beta$,
which leads to an ill-conditioned system and consequently convergence difficulties for any numerical solver. 
We addressed this by employing an incremental penalty approach, gradually
increasing the penalty parameters to guide the solution toward the feasible region. Such an approach
is strongly related to the interior point method (IPM) and Barrier methods. The name continuation
method is borrowed from topology optimization, where it is used to find the global minimum of a
non-convex problem by gradually increasing the penalty
parameter~\cite{rojas-labanda_automatic_2015}. In general, all these methods have in common to find
a global minimum of a non-convex minimization problem, which does not include friction. To mitigate
the risk of deviating from the local energy well, we propose a two-way penalty continuation. By
starting with the target penalty, gradually decreasing it until the system is well-conditioned
enough to find the solution, and finally increasing the penalty to the target again, we restrict the
nature of such an approach to land at the global minimum.
However, while this mitigates some of the numerical challenges, it undoubtedly introduces additional
complexity.

Another, obvious, challenge arises from the diffuse representation by incorporating the phase-field method.
The size of the transition zone, controlled by the phase-field energy gradient parameter $\epsilon$,
has a significant impact on the frictional behavior. A large transition zone softens the contact
and leads to a larger spread of the contact forces over a wider area, which may not be desirable in applications
requiring high precision. On the other hand, a small transition zone can lead to a high sensitivity to
mesh resolution and may introduce numerical artifacts~\cite{lorez_eulerian_2024}.
Furthermore, the nodes of a contact patch become volumetrically distributed rather than being confined to a single contact
line. This means that multiple layers of nodes within the contact patch contribute to the frictional
response. For the frictional force, this means that, while the forces integrated over the width of
the contact patch are consistent with the analytical Cattaneo-Mindlin solution
(see~\cref{fig:cattaneo-mindlin-problem}), the force density is not necessarily constant across the
contact patch. This may introduce unphysical shearing of the diffuse interface.

Looking forward, this work opens several avenues for future research.
Applying the method to model multiphysical systems involving contact and friction could provide
valuable insights into phenomena such as biofilm growth, where friction plays a significant role in
how the film grows~\cite{amar_patterns_2014}.
We believe our proposed framework can be helpful in understanding how growth processes are affected by frictional interaction with the environment,  
which could provide valuable insight in biology and materials science.

\section{Conclusion}
\label{sec:conclusion}

In this paper, we extended the novel Eulerian framework incorporating a diffuse phase-field
representation presented in~\cite{lorez_eulerian_2024} to frictional contact between soft bodies. 
We propose to include a frictional penalty force field as a function of Eulerian field variables,
namely the phase-fields $\phi$ and the velocities $\boldsymbol{v}$ of the bodies in contact.
Using multiple velocity fields renders the bodies independent, coupled only through contact forces
arising from frictional interactions.

We thoroughly validated the method with various numerical examples, demonstrating accuracy by comparison with the
analytical Cattaneo-Mindlin solution for frictional contact. Our examples also highlighted the
ability to capture history dependence and energy dissipation in frictional contact, respecting the
stick-slip transition and providing oscillation-free friction during sliding.
We believe that the proposed framework could be employed to complex multi-physical systems where interfacial interactions play a key role.

\section*{CRediT authorship contribution statement}
\textbf{Flavio Lorez:} Conceptualization, Methodology, Formal analysis, Investigation, Software, Writing - Original Draft, Visualization;
\textbf{Mohit Pundir:} Conceptualization, Writing - Review \& Editing, Supervision, Funding acquisition

\section*{Acknowledgments}
This work was supported by an ETH Research Grant (24-1 ETH-020).
MP acknowledges support from the Swiss National Science Foundation under the SNSF starting
grant (TMSGI2\_211655).
We would also like to thank David S. Kammer and Antoine Sanner for their valuable discussions and insights, which greatly contributed to the development of this work.

\section*{Data Availability}
The code and generated simulation data from this study have been deposited in the ETH Research
Collection database under accession code \href{https://bamboost.ch}{[url will be inserted during the proof process]}.

\end{document}